\documentclass[submission,copyright,creativecommons]{eptcs}
\providecommand{\event}{ICLP Doctoral Consortium 2020}
\usepackage{bytefield, rotating}

\title{A Low-Level Index for Distributed Logic Programming}
\author{Thomas Prokosch
  \institute{Institute for Informatics, Ludwig-Maximilian University
    of Munich, Germany}
  \email{prokosch@pms.ifi.lmu.de}
}

\def\titlerunning{A Low-Level Index for Distributed Logic Programming}
\def\authorrunning{Thomas Prokosch}
\begin{document}
\maketitle

\begin{abstract}
  A distributed logic programming language with support for
  meta-programming and stream processing offers a variety of
  interesting research problems, such as: How can a versatile and
  stable data structure for the indexing of a large number of
  expressions be implemented with simple low-level data structures?
  Can low-level programming help to reduce the number of occur checks
  in Robinson's unification algorithm? This article gives the answers.
\end{abstract}

\section{Introduction and problem description}
Logic programming originated in the 1970s as a result on work in
artificial intelligence and automated theorem
proving~\cite{green_logprog,kowalski_logprog}. One important concept
of logic programming always stood out: The clear separation between
the logic component and the control component of a
program~\cite{kowalski_alc}. In today's computing landscape, where
large amounts of (possibly streamed) data and distributed systems with
parallel processors are the norm, it becomes increasingly hard to
program in an imperative style where logic and control are
intermingled.

Therefore, it is worthwhile to investigate how a logic programming
language could deal with large amounts of streamed and non-streamed
data in a way such that it can adapt itself to changing circumstances
such as network outages (``meta-programming''). Creating such a
programming language is the primary drive behind the author's line of
research.

The main components of a distributed logic programming language are
\begin{itemize}
\item a stable indexing data structure to store large amounts of expressions,
\item a low-level unification algorithm with almost linear performance, and
\item a distributed forward-chaining resolution-based inference engine.
\end{itemize}

Some of these components have already been investigated; the current
status of the research is summarized in this article. The missing
parts are outlined in section~\ref{sec:open-issues-goals}.

\section{Logical foundations}
This section introduces standard algebraic terminology and is based
on~\cite{prokosch_unifrun,prokosch_reasoningtrie}.

Let $v_0, v_1, v_2, \ldots$ denote infinitely many variables, letters
$a, b, c, \ldots$ (except $v$) denote finitely many non-variable
symbols. $v^i$ (with a superscript) denotes an arbitrary variable.

An \emph{expression} is either a first-order term or a first-order
formula. Expressions are defined as follows: A variable is an
expression. A nullary expression constructor $c$ consisting of the
single non-variable symbol $c$ is an expression. If $e_1, \ldots, e_n$
are expressions then $c(e_1, \ldots, e_n)$ is an expression with
expression constructor $c$ and arity $n$.

The fusion of the two distinct entities term and formula may seem
unusual at first glance. This perspective, however, is convenient for
meta-programming: Meta-programming is concerned with the generation
and/or modification of program code through program code. Thus,
meta-programming applied to logic programming may require the
modification of formulas through functions which may be difficult to
achieve when there is a strict distinction between terms and formulas.
Commonly, a so-called \emph{quotation} is used to maintain such a
distinction when it is necessary to allow formulas to occur inside of
terms. However, it was
shown~\cite{chen_hilog,jiang_amblog,kalsbeek_vademecum} that it is not
necessary to preserve such a distinction and that by removing it, the
resulting language is a conservative extension of first-order
logic~\cite{bry_meta}.

Let $E$ denote the set of expressions and $V$ the set of variables. A
substitution $\sigma$ is a total function $V\to E$ of the form
$\sigma = \{v^1\mapsto e_1, \ldots, v^n\mapsto e_n\},\ n\geq 0$ such
that $v^1, \ldots, v^n$ are pairwise distinct variables, and
$\forall i\in\{1,\ldots,n\}$ $\sigma(v^i)=e_i$, and $\sigma(v)=v$ if
$v\neq v^i$. A substitution $\sigma$ is a renaming substitution iff
$\sigma$ is a permutation of variables, that is
$\{v^i\mid1\leq i\leq n\}=\{e_i\mid1\leq i\leq n\}$. $\sigma$ is a
renaming substitution for an expression $e$ iff
$\{e_i\mid1\leq i\leq n\}\subseteq V$ and for all distinct variables
$v^j, v^k$ in $e$ the inequality $\sigma(v^j)\neq\sigma(v^k)$ holds.

The application of a substitution $\sigma$ to an expression $e$,
written $\sigma(e)$, is defined as the usual function application,
i.e.\ all variables $v^i$ in $e$ are simultaneously substituted with
expressions $\sigma(v^i)$. The application of a substitution $\sigma$
to a substitution $\tau$, written $\sigma\tau$, is defined as
$(\sigma\tau)x=\sigma(\tau(x))$.

\section{Low-level representations}
One of the most important aspects in designing efficient algorithms is
finding a good in-memory representation of the key data structures.
The in-memory representation of variables, expressions, and
substitutions described in this section has already been published
in~\cite{prokosch_unifrun,prokosch_reasoningtrie} and is based on the
prefix notation of expressions. The prefix notation is a flat
representation without parentheses; the lack of parentheses makes this
representation especially suited for the flat memory address space of
common hardware. For example, the prefix notation of the expression
$f(a, v_1, g(b), v_2, v_2)$ is $f/5\ a/0\ v_1\ g/1\ b/0\ v_2\ v_2$.

A similar but distinct expression representation is the flatterm
representation~\cite{christian_kbc,christian_flat}.

\subsection{Representation of expressions}
An expression representation that is particularly suitable for a
run-time system of a logic programming language is as follows: Each
expression constructor is stored as a compound of its symbol $s$ and
its arity $n$. Each variable either stores the special value
\texttt{nil} if the variable is unbound or a pointer if the variable
is bound. It is worth stressing that the name of a variable is
irrelevant since its memory address is sufficient to uniquely identify
a variable. Two distinct expression representations do not share
variables.

In order to be able to represent non-linear expressions, i.e.\ 
expressions in which a variable occurs more than once, two kinds of
variables need to be distinguished: Non-offset variables and offset
variables. The first occurrence of a variable is always a non-offset
variable, represented as described above. All following occurrences of
this variable are offset variables and are represented by a pointer to
the memory address of the variable's first occurrence. Care must be
taken when setting the value of an offset variable: Not the memory
cell of the offset variable is modified but the memory cell of the
base variable it refers to.

The type of the memory cell (i.e.\ expression constructor
\texttt{cons}, non-offset variable \texttt{novar}, or offset variable
\texttt{ofvar}) is stored as a three-valued flag at the beginning of
the memory cell. Assuming that a memory cell has a size of 4 bytes, a
faithful representation of the expression $f(a, v_1, g(b), v_2, v_2)$
starting at memory address $0$ is:

\begin{center}
  \medskip
  \begin{bytefield}{28}
    \bitheader{0-27} \\
    \bitbox{1}{\rotatebox{90}{\tiny cons}}  &
    \bitbox{3}{$f/5$} &
    \bitbox{1}{\rotatebox{90}{\tiny cons}}  &
    \bitbox{3}{$a/0$} &
    \bitbox{1}{\rotatebox{90}{\tiny novar}} &
    \bitbox{3}{nil}   &
    \bitbox{1}{\rotatebox{90}{\tiny cons}}  &
    \bitbox{3}{$g/1$} &
    \bitbox{1}{\rotatebox{90}{\tiny cons}}  &
    \bitbox{3}{$b/0$} &
    \bitbox{1}{\rotatebox{90}{\tiny novar}} &
    \bitbox{3}{nil}   &
    \bitbox{1}{\rotatebox{90}{\tiny ofvar}} &
    \bitbox{3}{4}
  \end{bytefield}
\end{center}

The offset variable at address 24 contains the value 4 which must be
subtracted from its address yielding 20, the address of the base
variable the offset variable refers to.

Reading an in-memory expression representation involves traversing the
memory cells from left to right while keeping a counter of the number
of memory cells still to read. Each read memory cell decreases this
counter by one, and the arities of expression constructors are added
to the counter. Eventually, the counter will drop to zero which means
that the expression has been read in its entirety.

In subsequent examples the expression representation is simplified to
not include type flags.

\subsection{Representation of substitutions and substitution application}

An elementary substitution $\{v^i\mapsto e\}$ is represented as a
tuple of two memory addresses, the address of the variable $v^i$ and
the address of the first memory cell of the expression representation
of $e$. A substitution is represented as a list of tuples of
addresses. Assume the representation of the expression
$f(a, v_1, g(b), v_2, v_2)$ starts at address $0$ and the
representation of the expression $h(a, v_3)$ at address $36$, then the
substitution $\{v_2\mapsto h(a, v_3)\}$ is represented as the tuple
$(20, 36)$:

\begin{center}
  \medskip
  \begin{bytefield}{48}
    \bitheader{0-47} \\
    \bitbox{4}{$f/5$}    &
    \bitbox{4}{$a/0$}    &
    \bitbox{4}{nil}      &
    \bitbox{4}{$g/1$}    &
    \bitbox{4}{$b/0$}    &
    \bitbox{4}{nil}      &
    \bitbox{4}{4}        &
    \bitbox{8}{$\ldots$} &
    \bitbox{4}{$h/2$}    &
    \bitbox{4}{$a/0$}    &
    \bitbox{4}{$nil$}    &
  \end{bytefield}
\end{center}

Substitution application simply consists of setting the contents of
the memory cell of the variable to the address of the expression
representation to be substituted. After the substitution application
$f(a, v_1, g(b), v_2, v_2)\{v_2\mapsto h(a, v_3)\}$ memory contents
is:

\begin{center}
  \medskip
  \begin{bytefield}{48}
    \bitheader{0-47} \\
    \bitbox{4}{$f/5$}    &
    \bitbox{4}{$a/0$}    &
    \bitbox{4}{nil}      &
    \bitbox{4}{$g/1$}    &
    \bitbox{4}{$b/0$}    &
    \bitbox{4}{$36$}     &
    \bitbox{4}{4}        &
    \bitbox{8}{$\ldots$} &
    \bitbox{4}{$h/2$}    &
    \bitbox{4}{$a/0$}    &
    \bitbox{4}{$nil$}    &
  \end{bytefield}
\end{center}

Observe that the contents of the offset variable at address $24$
keeps its offset $4$ unchanged, and that the contents of the
non-offset variable at address $20$ contains an absolute address.

\section{Storage and retrieval}
Automated reasoning~\cite{robinson_har} relies upon the efficient
storage and retrieval of expressions. Standard data structures such as
lists or hash tables can be used for this task but more specialized
data structures, known as term
indexing~\cite{graf_termidx,sekar_termidx} data structures, can
significantly improve the retrieval speed of
expressions~\cite{christian_flat,sekar_termidx,schulz_pyres}.
Depending on the application, certain characteristics of a term
indexing data structure are beneficial. For
meta-programming~\cite{bry_meta} the retrieval of expressions
unifiable with a query as well as retrieval of instances,
generalizations, and variants of a query are desirable. For
tabling~\cite{tamaki_oldt,ramakrishnan_bktrk,johnson_tst}, a
form of memoing used in logic programming, the retrieval of variants
and generalizations of queries needs to be well supported. In this
section, which is based upon already published
work~\cite{prokosch_reasoningtrie}, \emph{instance tries} are
proposed. Instance tries are trees which offer a few conspicuous
properties such as:
\begin{itemize}
\item Stability. Instance tries are stable in the sense that the order
  of insertions into and removals from the data structure does not
  determine its shape.
\item Versatility. Instance tries support the retrieval of
  generalizations, variants, and instances of a query as well as those
  expressions unifiable with a query.
\item Incrementality. Instance tries are based upon the instance
  relationship which allows for incremental unification during
  expression retrieval.
\item Perfect filtering. Some term indexing data structures require
  that the results of a query are post-processed~\cite{graf_termidx}.
  Instance tries do not require such post-processing because querying
  an instance trie always returns perfect results.
\end{itemize}

\subsection{Review of related work}
Term indexing data structures are surveyed in the
book~\cite{graf_termidx} and in the book chapter~\cite{sekar_termidx}
the latter containing some additional data structures which did not
exist when the former was written. The latter does not describe dated
term indexing data structures.

Tries were invented in 1959 for information
re\emph{trie}val~\cite{briandais_trie} while the name itself was
coined one year later~\cite{fredkin_trie}. Tries exhibit a more
conservative memory usage than lists or hash tables due to the fact
that common word prefixes are shared and thus stored only once.

Coordinate indexing~\cite{hewitt_planner} and path
indexing~\cite{stickel_pathidx} consider positions (or sequences of
positions, respectively, so-called \emph{paths}) of symbols in a term
with the goal of subdividing the set of terms into subsets. Both
coordinate indexing and path indexing disregard variables in order to
further lower memory consumption making them non-perfect filters:
Subject to these limitations, terms $f(v_0, v_1)$ and $f(v_0, v_0)$
are considered to be equal which means that results returned from a
query need to be post-processed to identify the false positives.
Several variations of path indexing, such as Dynamic Path
Indexing~\cite{letz_setheo} and Extended Path
Indexing~\cite{graf_extpathidx} have been proposed, none of which are
stable or perfect filters.

Discrimination trees~\cite{mccune_discrimtrees,mccune_exptermidx}
(with their variants Deterministic Discrimination
Trees~\cite{graef_ltrtrpat} and Adaptive Discrimination
Trees~\cite{sekar_apm}) were proposed as the first tree data
structures particularly designed for term storage. However, all of
them are non-perfect filters, a shortcoming that Abstraction
Trees~\cite{ohlbach_abstree} were able to remedy. Substitution
Trees~\cite{graf_substreeidx} and Downward Substitution
Trees~\cite{hoder_unifcmp} further refine the idea of abstraction
trees and have been recently extended to also support indexing of
higher order terms~\cite{pientka_hoti}.

While Code Trees~\cite{voronkov_avampire} and Coded Context
Trees~\cite{ganzinger_ctxtr} are also frequently used in automated
theorem provers both data structures are not versatile according to
the characterization above.

\subsection{Order on expressions}
A total order on expressions $\leq_e$ is lexicographically derived
from the total order $\leq_c$:
\begin{itemize}
\item $v_1<_cv_2$ for variables $v_1, v_2$ with an order $\leq_v$ such
  that $v_1<_vv_2$,
\item $v<_cs/a$ for variable $v$ and non-variable symbol $s$ with
  arity $a$,
\item $s/a_1<_cs/a_2$ for non-variable symbol $s$ and $a_1<a_2$,
\item $s_1/a_1<_cs_2/a_2$ for non-variable symbols $s_1, s_2$ with an
  order $\leq_{nv}$ such that $s_1<_{nv}s_2$.
\end{itemize}

\subsection{Matching and unification modes}
A versatile term indexing data structure needs to support the
retrieval of expressions that are more general than, a variant of, an
instance of, or unifiable with a query. For the construction and
querying of instance tries, however, mutually exclusive definitions
are required. Expression $e_1$ is a variant (VR) of expression $e_2$
iff there exists a renaming substitution $\rho$ for $e_1$ such that
$e_1\rho=e_2$. $e_1$ is strictly more general (SG) than $e_2$ and
$e_2$ is a strict instance (SI) of $e_1$ iff $\sigma$ is a
non-renaming substitution for $e_1$ such that $e_1\sigma=e_2$. $e_1$
and $e_2$ are only unifiable (OU) iff
$\exists\sigma\ (e_1\sigma=e_2\sigma)$, $\sigma$ is most general, and
$\sigma$ is not a renaming substitution for both $e_1$ and $e_2$.
$e_1$ and $e_2$ are non-unifiable (NU) iff
$\forall\sigma\ (e_1\sigma\ne e_2\sigma)$. A matching-unification
algorithm that is able to determine the mode for two expressions is
given in Section~\ref{sec:low-level-unif}.

\subsection{Instance tries}
An instance trie is a tree $T$ such that
\begin{itemize}
\item Every node of $T$ except the root stores an expression.
\item Every child $N_c$ of a node $N$ in $T$ is a strict instance of $N$.
\item Siblings $N_s$ in $T$ are ordered by $<_e$ as described above.
\end{itemize}

\begin{figure}
  \centering
  \begin{minipage}[b]{0.40\textwidth}
    \centering
    \includegraphics[width=0.9\textwidth]{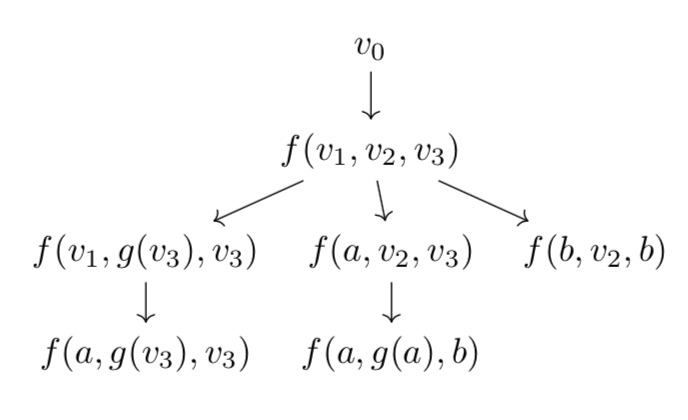}
    \caption{Example of an instance trie}\label{fig:inst1}
  \end{minipage}\hfill
  \begin{minipage}[b]{0.50\textwidth}
    \centering
    \includegraphics[width=0.9\textwidth]{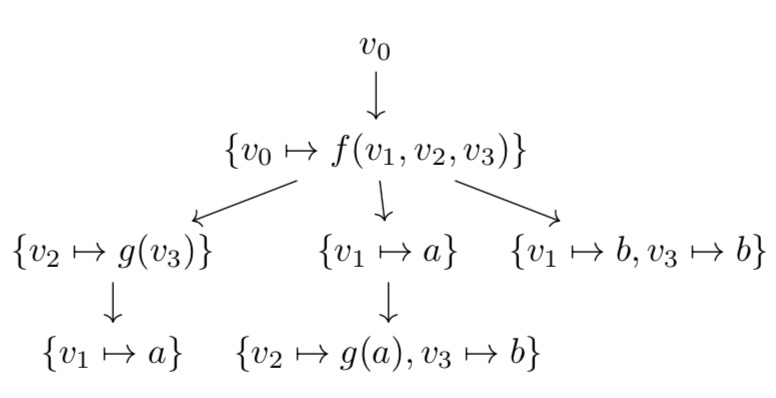}
    \caption{Instance trie using substitutions}\label{fig:inst2}
  \end{minipage}
\end{figure}

Figure \ref{fig:inst1} shows an example of an instance trie storing
six expressions. Figure \ref{fig:inst2} shows the same trie using
substitutions; this alternate representation is possible due to the
strict instance relation between a node and its children. In this
representation, repeated application of substitutions to the root
variable $v_0$ along a path yields the corresponding expression. The
use of substitutions gives two advantages: First, common symbols are
shared further reducing memory consumption. Second, querying an
instance trie can make use of incremental unification resulting in
faster retrieval.

\emph{Retrieving} expressions from an instance trie $T$ requires the
following steps:
\begin{enumerate}
\item Top-down left-to-right traversal of the tree.
\item Expression $e$ of each node $N$ is unified with the query $q$ to
  determine the matching and unification mode of $e$ and $q$ as
  outlined above.
\end{enumerate}
All query modes except unification affect the traversal:
\begin{itemize}
\item Variant: The traversal can be interrupted as soon as an answer
  is found.
\item Instance: If $e$ is an instance of $q$ then the sub-tree rooted
  at $N$ does not need to be traversed: All children of $N$
  necessarily store instances of $q$.
\item Generalization: The traversal can ignore all child nodes of a
  node which is not strictly more general than $q$.
\end{itemize}

\emph{Insertion} of an expression $e$ into an instance trie $T$
involves first searching for a node $N$ such that the expression of
$N$ is more general than $e$ and the expressions of all children of
$N$ are not more general than $e$. If $e$ is a variant of the
expression in $N$ then nothing is done. Otherwise, a new node $N'$
containing expression $e$ is inserted as a child of $N$ (at the
correct position among its siblings according to the order $<_e$
defined above) and instances of $e$ (found to the right of $N'$) are
inserted below $N'$.

\emph{Deletion} of an expression $e$ from an instance trie $T$
requires retrieving the node $N$ containing expression $e$. If such a
node is found then this node $N$ is deleted and, after this deletion,
each child node of $N$ is inserted into the node which, before the
deletion, was the parent node of $N$.

\section{Low-level unification}
\label{sec:low-level-unif}
Unification, that is determining whether a pair of expressions has a
most general unifier (MGU), is an integral part of every automated
reasoning system and every logic programming language. Nevertheless,
only little attention has been given to potential improvements which
develop their full effect at machine level or in an interpreter
run-time. This section, based on previously published
work~\cite{prokosch_unifrun}, outlines a unification algorithm which
has been specifically developed for such an environment.

\subsection{Review of related work}
Since Robinson introduced unification~\cite{robinson_resolution}, a
wealth of research has been carried out on this
subject~\cite{paterson_linunif, champeaux_pwunif, martelli_effunif,
  escalada_linunif}. Nevertheless, only few algorithms are used in
practice not only because more sophisticated algorithms are harder to
implement but also, unexpectedly, Robinson's unification algorithm is
still the most efficient~\cite{hoder_unifcmp}! Consequently,
Robinson's unification algorithm has been chosen as a starting point
for the following unification algorithm.

\subsection{A matching-unification algorithm}
The algorithm \texttt{unif(e1, e2)} performs a left-to-right traversal
of the representation of expressions $e_1$ and $e_2$ whose first
addresses are \texttt{e1} and \texttt{e2}, respectively. Let
\texttt{c}, \texttt{c1}, \texttt{c2} be addresses of memory cells in
the representation of $e_1$ or $e_2$. In each step of the algorithm
two memory cells are compared based on type and content using the
following functions:
\begin{itemize}
\item \texttt{type(c)}: Returns the type of the value stored at
  \texttt{c}, resulting in \texttt{cons}, \texttt{novar}, or
  \texttt{ofvar}.
\item \texttt{value(c)}: Value stored in memory cell \texttt{c}.
\item \texttt{arity(c)}: Arity of the constructor stored in memory
  cell \texttt{c}, or $0$.
\item \texttt{deref(c, S)}: Creates a new expression from the
  expression representation at \texttt{c} and applies substitution
  \texttt{S} to it.
\item \texttt{occurs-in(c1, c2)}: Checks whether variable at
  \texttt{c1} occurs in expression at \texttt{c2}.
\end{itemize}
The algorithm sets and uses the following variables: \texttt{A} (short
for ``answer'') is initialized with \texttt{VR} and contains
\texttt{VR}, \texttt{SG}, \texttt{SI}, \texttt{OU}, or \texttt{NU}.
Variables \texttt{R1} and \texttt{R2}, both initialized with 1,
contain the number of remaining memory cells to read. The algorithm
terminates if \texttt{R1 = 0} and \texttt{R2 = 0}. \texttt{S1} and
\texttt{S2} contain substitutions for variables in the expression
representations of $e_1, e_2$ and are initialized with the empty lists
\texttt{S1 := []}, \texttt{S2 := []}.

In each step of the algorithm two memory cells \texttt{c1} and
\texttt{c2} are compared (starting with \texttt{e1} and \texttt{e2},
respectively), with the following four possibilities for each cell,
resulting in a total of 16 cases: \texttt{type(ei) = cons},
\texttt{type(ei) = novar}, \texttt{type(ei) = ofvar \&\& deref(ei, Si)
  != nil}, and \texttt{type(ei) = ofvar \&\& deref(ei, Si) = nil}.

Table \ref{tab:matchunif} shows the core of the matching-unification
algorithm. For clarity and because the table is symmetric along its
principal diagonal only the top-right half of the table contains
entries. For space reasons, the table is abbreviated; for the full
table refer to~\cite{prokosch_unifrun}.

\begin{table}\centering
  \begin{tabular}{l|llll}
    case                 & \texttt{cons}  & \texttt{novar} & \texttt{ofvar \&\& deref!=nil} & \texttt{ofvar \&\& deref=nil} \\
    \hline
    \texttt{cons}        & continue or NU & bind,          & dereference                    & \emph{occurs check}           \\
                         &                & change mode    & recursive call                 & (bind and OU) or NU           \\
    \hline
    \texttt{novar}       &                & bind to left   & bind, change mode              & bind, change mode             \\
    \hline
    \texttt{ofvar \&\&}  &                &                & dereference                    & \emph{occurs check}           \\
    \texttt{deref!=nil}  &                &                & recursive call                 & (bind and OU) or NU           \\
    \hline
    \texttt{ofvar \&\&}  &                &                &                                & bind                          \\
    \texttt{deref\ =nil} &                &                &                                &                               \\
  \end{tabular}
  \caption{Core of the matching-unification algorithm, abbreviated}
  \label{tab:matchunif}
\end{table}

\subsection{Illustration of the matching-unification algorithm}
An example should illustrate how the matching-unification algorithm
works: Expression $f(v_1, v_1)$ at address \texttt{e1=0} shall be
unified with expression $f(a, a)$ at address \texttt{e2=20}:

\begin{center}
  \medskip
  \begin{bytefield}{32}
    \bitheader{0-31}    \\
    \bitbox{4}{$f/2$}    &
    \bitbox{4}{nil}      &
    \bitbox{4}{$4$}      &
    \bitbox{8}{$\ldots$} &
    \bitbox{4}{$f/2$}    &
    \bitbox{4}{$a/0$}    &
    \bitbox{4}{$a/0$}
  \end{bytefield}
\end{center}

\begin{enumerate}
\item Initialization: \texttt{A := VR; R1 := 1; R2 := 1; S1 := []; S2 := []}
\item \texttt{type(e1) = cons, type(e2) = cons, value(e1) = value(e2)}

  Constructors $f/2$ and $f/2$ match with arity 2: \texttt{R1 := R1+2
    = 3; R2 := R2+2 = 3}
\item Continue to next cell: Each cell consists of 4 bytes.

  \texttt{e1 := e1+4 = 4; e2 := e2+4 = 24; R1 := R1-1 = 2; R2 := R2-1 = 2}
\item \texttt{type(e1) = novar, type(e2) = cons}

  First, the non-offset variable at address \texttt{e1=4} needs to be
  bound to the sub-expression starting with the constructor $a/0$ at
  address \texttt{e2=24} by adding the tuple $(4, 24)$ to the
  substitution \texttt{S1 := [(4, 24)]}. Note that no occurs check is
  required when introducing this binding; this is a speed improvement
  with respect to some other unification algorithms such as Robinson's
  algorithm~\cite{robinson_resolution} or the algorithm from
  Martelli-Montanari~\cite{martelli_effunif}.

  Then, change the mode by setting \texttt{A:=SG} since the non-offset
  variable at \texttt{e1=4} is strictly more general than the
  expression constructor $a/0$ at \texttt{e2=24}.
\item Continue to next cell: Each cell consists of 4 bytes.

  \texttt{e1 := e1+4 = 8 ; e2 := e2+4 = 28; R1 := R1-1 = 1; R2 := R2-1 = 1}
\item \texttt{type(e1) = ofvar, type(e2) = cons}

  First, dereference \texttt{e1} with \texttt{S1} yielding address 24.
  Dereferencing address \texttt{e2=28} yields 28. (The memory cell at
  address 28 contains a constructor.) Then, call the algorithm
  recursively with addresses 24 and 28.
\item The recursive call confirms the equality of the expression
  constructors $a/0$ at address 24 and $a/0$ at address 28, returning
  to the caller without any changes to variables \texttt{A},
  \texttt{R1}, \texttt{R2}.
\item Continue to next cell: Each cell consists of 4 bytes.

  \texttt{e1 := e1+4 = 12 ; e2 := e2+4 = 32; R1 := R1-1 = 0; R2 := R2-1 = 0}

  The algorithm terminates because \texttt{R1 = 0} and \texttt{R2 =
    0}. The result \texttt{A = SG} and \texttt{S1 = [(4, 24)]},
  \texttt{S2 = []} is returned to the caller. The result is correct
  since $f(v_1, v_1)$ is strictly more general than $f(a, a)$.
\end{enumerate}

\section{Open issues and goals}
\label{sec:open-issues-goals}
While some progress towards a distributed logic programming language
has already been made, there are still further challenges:

\begin{itemize}
\item Instance tries have been fully specified and their
  implementation is currently ongoing. Upon completion an empirical
  evaluation of instance tries together with a variety of common term
  indexing data structures need to verify the expected speed-up of
  instance tries.

\item It is currently being investigated how stream processing can be
  integrated with logic programming.

\item The forward-chaining resolution engine to derive the immediate
  consequences from a set of expressions in order to perform program
  evaluation has not been investigated so far.
\end{itemize}

Follow-up articles will report on each of those aspects of research.


\end{document}